\documentclass[english,aip,apl,reprint]{revtex4}
\usepackage[T1]{fontenc}
\usepackage[latin9]{inputenc}
\usepackage{color}
\definecolor{document_fontcolor}{rgb}{0, 0, 0}
\color{document_fontcolor}
\usepackage{babel}
\usepackage{float}
\usepackage{bm}
\usepackage{amssymb}
\usepackage{graphicx}
\usepackage[unicode=true,pdfusetitle,
 bookmarks=true,bookmarksnumbered=false,bookmarksopen=false,
 breaklinks=false,pdfborder={0 0 1},backref=false,colorlinks=true]
 {hyperref}

\makeatletter
\@ifundefined{textcolor}{}
{%
 \definecolor{BLACK}{gray}{0}
 \definecolor{WHITE}{gray}{1}
 \definecolor{RED}{rgb}{1,0,0}
 \definecolor{GREEN}{rgb}{0,1,0}
 \definecolor{BLUE}{rgb}{0,0,1}
 \definecolor{CYAN}{cmyk}{1,0,0,0}
 \definecolor{MAGENTA}{cmyk}{0,1,0,0}
 \definecolor{YELLOW}{cmyk}{0,0,1,0}
}

\usepackage{babel}

\makeatother

\begin{document}
\title{Surface-induced reduction of the switching field in nanomagnets }
\author{R. Bastardis and H. Kachkachi}
\email{hamid.kachkachi@univ-perp.fr}

\affiliation{Laboratoire PROMES CNRS UPR8521, Université de Perpignan Via Domitia,
Rambla de la thermodynamique, Tecnosud, F-66100 Perpignan, France}
\date{\today}
\begin{abstract}
Magnetization reversal in a many-spin nanomagnet subjected to an rf
magnetic field, on top of a DC magnetic field, is studied by numerically
solving the system of coupled (damped) Landau-Lifshitz equations.
It is demonstrated that spin-misalignment induced by surface anisotropy
favors switching with a DC magnetic field weaker than the Stoner-Wohlfarth
switching field, for optimal intensities and frequencies of the rf
field. 
\end{abstract}
\maketitle

\section{introduction}

Storing data on magnetic media involves magnetization reversal between
two magnetic states, the $1$ and $0$ states. In recent years, in
order to optimize magnetic recording, a variety of nanoscale magnetic
systems has been extensively investigated. However, the two magnetic
states in such systems are usually separated by high magnetic anisotropy
energy barriers. While this is favorable to thermal stability of the
data stored, high magnetic fields are required for their switching
(and thus for writing and reading). The problem is that these fields
are rather difficult to achieve at the nanoscale because of heating
and noise issues, for instance. As such, several strategies have been
attempted to overcome such a hurdle, and one of which consists in
applying a time-dependent magnetic (rf) field. This has been shown
to considerably reduce the static switching field (the Stoner-Wohlfarth
switching field), see \emph{e.g.} Refs. \citep{thirionetal03nat,woltersdorfetal07prl,barros11prb,barrosetal13prb,SutoEtAl_apl2017}
and many references therein. Several works \citep{SunWang2006,barrosetal13prb,CaiEtAl_prb13,TaniguchiEtAl_prb16}
have proceeded by applying an rf field and solving the Landau-Lifshitz
equation in the macrospin approach. Alternatively, it was shown in
Refs. \citep{garkacrey08epl,garkac09prb} that the magnetization switching
in nanomagnets in the many-spin approach may be achieved via internal
second-generation spin-wave modes. Recent experimental and theoretical
studies \citep{SekiEtAl_nc2013,YamajiImamura_apl2016} have also investigated
switching assisted by spin waves and have showed that these lead to
a reduction of the switching field.

In general, the generation of spin waves is triggered by spin misalignment
and, in nanomagnets, the latter is most often caused by boundary effects
and surface anisotropy. Accordingly, in the present work, we investigate
how surface anisotropy affects switching and, in particular, how the
switching field intensity and frequency depend on surface anisotropy
and the size of the nanomagnet. For this purpose, we consider the
simplest nanomagnet of cubic shape, an underlying simple cubic lattice
structure, and a typical set of energy parameters such as exchange
coupling, core and surface anisotropy constants \citep{kacgar05springer,SchmoolKachkachi_acapress2015,SchmoolKachkachi_acapress2016,KansoEtAl_prb19,IglesiasKachkachi_spring2021}.
We consider a nanomagnet of $\mathcal{N}=N_{x}\times N_{y}\times N_{z}$
classical (atomic) magnetic moments $\bm{m}_{i}=\mu_{a}\bm{s}_{i}$,
with $\bm{s}_{i}$ a unit vector and $\mu_{a}$ the amplitude. The
magnetic properties of such a nanomagnet may be investigated with
the help of the Dirac-Heisenberg Hamiltonian\citep{dimwys94prb,igllab01prb,kacdim02prb,kazantsevaetal08prb,fraileetal10prl,bastardiset12prb}
\begin{equation}
\mathcal{H}=-J\sum_{i,j}\mathbf{s}_{i}\cdot\mathbf{s}_{j}-\mathbf{h}\left(\tau\right)\cdot{\displaystyle {\displaystyle \sum_{i=1}^{\mathcal{N}}}}\mathbf{s}_{i}+{\displaystyle {\displaystyle \sum_{i=1}^{\mathcal{N}}}}\mathcal{H}_{{\rm an},i},\label{eq:HD-Ham}
\end{equation}
where the first term is the exchange energy with coupling $J$. In
general, one may introduce an exchange coupling that depends on the
loci of the spins. However, for the present study and the message
it aims to deliver, it is sufficient to consider a uniform exchange
coupling $J$ throughout the nanomagnet and use it as a reference
energy scale for the whole study, which means that all other parameters
are measured with respect to $J$. In the second term in Eq. (\ref{eq:HD-Ham}),
which is the Zeeman coupling energy, we introduced the reduced magnetic
field

\begin{equation}
\mathbf{h}\left(\tau\right)=\mathbf{h}_{\mathrm{dc}}+\mathbf{h}_{\mathrm{rf}},\label{eq:TotalMagneticField}
\end{equation}
comprising the static (or DC) magnetic field $\mathbf{h}_{\mathrm{dc}}=\left(\mu_{a}\mathbf{H}_{\mathrm{dc}}\right)/J$
and the time-dependent rf field $\mathbf{h}_{\mathrm{rf}}=\left(\mu_{a}\mathbf{H}_{\mathrm{rf}}\right)/J$.
The latter is taken circularly polarized in the $xy$ plane \citep{denisovetal06prl}
\begin{equation}
\mathbf{H}_{\mathrm{rf}}\left(\tau\right)=H_{{\rm rf}}^{0}\left[\cos\left(\varpi\tau\right)\mathbf{e}_{x}+\sin\left(\varpi\tau\right)\mathbf{e}_{y}\right]\label{eq:rfField}
\end{equation}
where $\tau\equiv t/\tau_{s}$ is the dimensionless time measured
with respect to the characteristic scaling time of the system's dynamics,
namely $\tau_{s}=\mu_{a}/\left(\gamma J\right)$, $\gamma$ being
the gyromagnetic ratio ($\sim$$1.76\times10^{11}\,{\rm T}^{-1}{\rm s}^{-1}$);
$\varpi=\omega\tau_{s}$ is the dimensionless frequency.

The last term in Eq. (\ref{eq:HD-Ham}) is the anisotropy energy contribution
with $\mathcal{H}_{{\rm an},i}$ being the on-site anisotropy energy.
Here, we distinguish between core and surface spins according to whether
they have their full coordination number (here $z=6$) or not. Indeed,
the nanomagnet under study is a box-shaped many-spin system with free
boundary conditions, with an underlying simple-cubic lattice. Therefore,
there are five different local environments with a coordination number
corresponding respectively to the core ($z=6$), facets ($z=5$),
edges ($z=4$) and corners ($z=3$). Then, to core spins we attribute
a uniaxial (on-site) anisotropy with easy axis along the $z$ direction
and constant $k_{c}=K_{c}/J>0$. On the other hand, the anisotropy
of surface spins is given by Néel's model \citep{nee54jpr}, according
to which anisotropy only occurs at sites with reduced coordination.
The corresponding anisotropy constant is denoted by $k_{s}=K_{s}/J>0$.
Therefore, we write

\begin{equation}
\mathcal{H}_{{\rm an},i}=\left\{ \begin{array}{ll}
-k_{c}\left(\mathbf{s}_{i}\cdot\mathbf{e}_{z}\right)^{2}, & i\in\mathrm{core}\\
\\
+\frac{1}{2}k_{s}{\displaystyle \sum_{j\in\mathrm{nn}}^{z_{i}}}\left(\mathbf{s}_{i}\cdot\mathbf{u}_{ij}\right)^{2}, & i\in\mathrm{surface}.
\end{array}\right.\label{eq:HamUA}
\end{equation}
$\mathbf{u}_{ij}$ is a unit vector connecting the nearest neighbors
(nn) $i$ and $j$.

\emph{Physical parameters}: A word is in order regarding the orders
of magnitude of these parameters. For cobalt, the lattice parameter
is $a=0.3554$ nm and the magnetic moment per atom is $\mu_{a}=n_{0}\mu_{B}$,
with $n_{0}$ being the number of Bohr magnetons per atom ($n_{0}\simeq1.7$)
and $\mu_{B}=9.274\times10^{-24}$J/T the Bohr magneton. Hence, $\mu_{a}\simeq1.58\times10^{-23}\,{\rm J/T}$.
The (bulk) exchange coupling is $J\simeq8\,\mathrm{mev}$ or $1.2834\times10^{-21}$
J/atom, which yields $\tau_{s}\sim70\,{\rm fs}$. Next, the magneto-crystalline
anisotropy constant is roughly $K_{c}\simeq3\times10^{-24}$ J/atom
and the surface anisotropy constant is $K_{s}\simeq5.22\times10^{-23}\,$
Joule/atom. As such, $k_{c}\equiv K_{c}/J\simeq0.00234$ and $k_{s}\equiv K_{s}/J\simeq0.04$.
The latter value is within the range of those inferred from several
experimental studies. Indeed, one may find $K_{s}/J\simeq0.1$ for
cobalt \citep{skocoe99iop}, $K_{s}/J\simeq0.06$ for iron \citep{urquhartetal88jap},
and $K_{s}/J\simeq0.04$ for maghemite particles \citep{perrai05springer}.
For later reference, we note that for a nanomagnet with the atomic
magnetic moment $\mu_{a}$ and magneto-crystalline anisotropy $K_{c}$
given above, the Stoner-Wohlfarth switching field $H_{{\rm SW}}=2K_{c}/\mu_{a}$
evaluates to $0.38\,{\rm T}$.

Another word is in order regarding the notations for the anisotropy
constants. The capital letter $K$ stands for the macroscopic constant
usually given in ${\rm J/}{\rm m}{}^{3}$ for the volume and ${\rm J/}{\rm m}{}^{2}$
for the surface. It can also be converted into ${\rm J/}{\rm atom}$
upon using the unit cell of the underlying lattice. The lowercase
letter $k$ is the dimensionless anisotropy constant that results
from a normalization with respect to the largest energy in the system,
namely the exchange coupling $J$, \emph{i.e. }$k\equiv K/J$, for
core and surface. In the figures presented below, instead of using
the dimensionless constant $k$, we use the anisotropy constant in
units of ${\rm meV}$ which should be understood as ${\rm meV}/{\rm atom}$.

The dynamics of the atomic spins $\bm{m}_{i}$ is described with a
set of ($2\mathcal{N}$) coupled damped Landau-Lifshitz equations
(LLE)

\begin{equation}
\frac{d\mathbf{s}_{i}}{d\tau}=-\mathbf{s}_{i}\times\bm{h}_{{\rm eff},i}-\alpha\,\mathbf{s}_{i}\times\left(\mathbf{s}_{i}\times\bm{h}_{{\rm eff},i}\right),\label{eq:LLE}
\end{equation}
where $\bm{h}_{{\rm eff},i}\equiv\left(\mu_{a}\mathbf{H}_{\mathrm{eff}}\right)/J$
is the (dimensionless) effective field acting on the atomic spin at
site $i$, given by $\bm{h}_{{\rm eff},i}=-\delta\mathcal{H}/\delta\bm{s}_{i}$;
$\alpha$ is the dimensionless damping parameter ($\sim0.01$). 

We use the (second-order) Heun algorithm to solve the set of equations
(\ref{eq:LLE}) starting from a given initial state (a configuration
of $\mathcal{N}$ spins $\bm{s}_{i}$) and a set of parameters $k_{c},k_{s},h_{{\rm dc}}$
and rf field characteristics, \emph{i.e. }amplitude $h_{{\rm rf}}=\mu_{a}H_{{\rm rf}}^{0}/J$
and frequency $\varpi=\omega\tau_{s}$. The calculations will be performed
for two anisotropy configurations: 
\begin{enumerate}
\item Model \textbf{TUA} (textured uniaxial anisotropy): all spins, core
and surface, have the same uniaxial anisotropy in the $z$ direction,
but the corresponding constant may be different for core and surface
spins. 
\item Model \textbf{NSA} (Néel surface anisotropy): the spins in the core
are attributed the uniaxial anisotropy easy axis according to the
first line in Eq. (\ref{eq:HamUA}) with constant $k_{c}$ and easy
axis $\bm{e}_{z}$, whereas those on the surface have their anisotropy
given by the second line with constant $k_{s}$. 
\end{enumerate}
In this study we vary the amplitude $h_{{\rm rf}}$ and frequency
$\varpi$ of the rf field and determine their optimal values for which
the net magnetic moment switches, for a given and fixed DC magnetic
field.

\emph{Numerical procedure}: we recall that the main objective here
is to achieve magnetization reversal under an rf field, in addition
to a DC magnetic field whose intensity is supposedly smaller than
the critical value required for switching according to the macrospin
Stoner-Wohlfarth model. In addition, we intend to demonstrate that
the intensity and frequency of the rf field that are usually required
by the switching of a macrospin, are further optimized in nanomagnets
which exhibit spin misalignment induced by surface anisotropy. Accordingly,
we prepare the system in an initial state, namely a spin configuration
with all spins aligned in a given direction, say in the $+z$ direction
and apply a DC magnetic field in the $-z$ direction. In fact, to
avoid having a vanishing vector product $\mathbf{s}_{i}\times\bm{H}_{{\rm dc}}$
at the initial time, we set $\bm{H}_{{\rm dc}}$ at an angle slightly
different from $\pi$ (\emph{e.g. }$179{^{\circ}}$). While maintaining
the rf magnetic field off, we set the damping parameter to some (relatively)
large value ($\sim1$) and let the spins evolve in time according
to Eq. (\ref{eq:LLE}) until the equilibrium state is reached. Since
the DC field is smaller than the switching field, the spins remain
in the local minimum defined by the anisotropy energy. Next, we reset
the damping parameter to $\alpha\sim0.02$ and switch on the rf field.
Then, we let the spins evolve in time, using again Eq. (\ref{eq:LLE}),
and record the net magnetic moment $\bm{m}\left(t\right)$ defined
as

\begin{equation}
\bm{m}=\frac{1}{\mathcal{N}}\sum_{i=1}^{\mathcal{N}}\mathbf{m}_{i}.\label{eq:Macrospin}
\end{equation}

\begin{figure}[H]
\centering{}\includegraphics[scale=0.215]{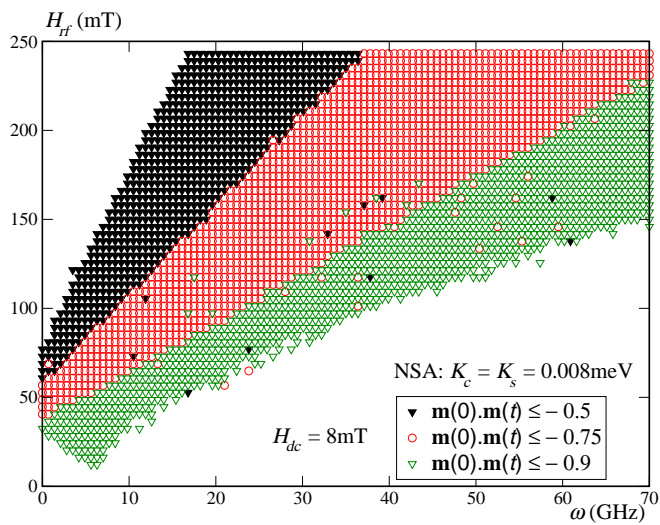}
\protect\caption{\label{fig:SwitchingCondition}Switching phase diagram for different
switching conditions. Each symbol represents a point of the diagram
for which switching is achieved. $\mathcal{N}=11^{3}$.}
\end{figure}

We stop this time evolution when two criteria are satisfied: First
we require that $\bm{m}\left(0\right)\cdot\bm{m}\left(t\right)\leq\varepsilon$,
with $\varepsilon$ a negative number. This condition is required
to ensure switching. Accordingly, in Fig. \ref{fig:SwitchingCondition},
we show the $\varpi$-$h_{{\rm rf}}$ phase diagram for the NSA model
with different values of $\varepsilon$ ($\left|\varepsilon\right|<1$).
We see that indeed the phase diagram strongly depends on the switching
criterion. More explicitly, for a given $h_{{\rm rf}}$, a higher
$\left|\varepsilon\right|$ requires a higher frequency $\varpi$
since more energy has to be pumped into the system from the rf field.
On the other hand, at a given frequency $\varpi$, the larger $\left|\varepsilon\right|$
the smaller is the critical $h_{{\rm rf}}$. This can be understood
by the fact that a small value of $\left|\varepsilon\right|$ requires
the system to be maintained halfway from the more stable (deeper)
minimum and this requires a stronger field. In all subsequent results
of the present study, we adopted the most stringent criterion, \emph{i.e.}
with the largest value of $\left|\varepsilon\right|$. At finite temperature,
this would insure the highest thermal stability for the magnetic state
of the system. In fact, a second criterion is necessary in order to
ensure that the final magnetic moment remains in the minimum reached
after switching. For this, we require that the fluctuations of the
net magnetic moment decay to a small value, \emph{i.e.} $\left\Vert \delta\bm{m}\left(t\right)\right\Vert \lesssim10^{-3}$.

\begin{figure}
\centering{}\includegraphics[scale=0.25]{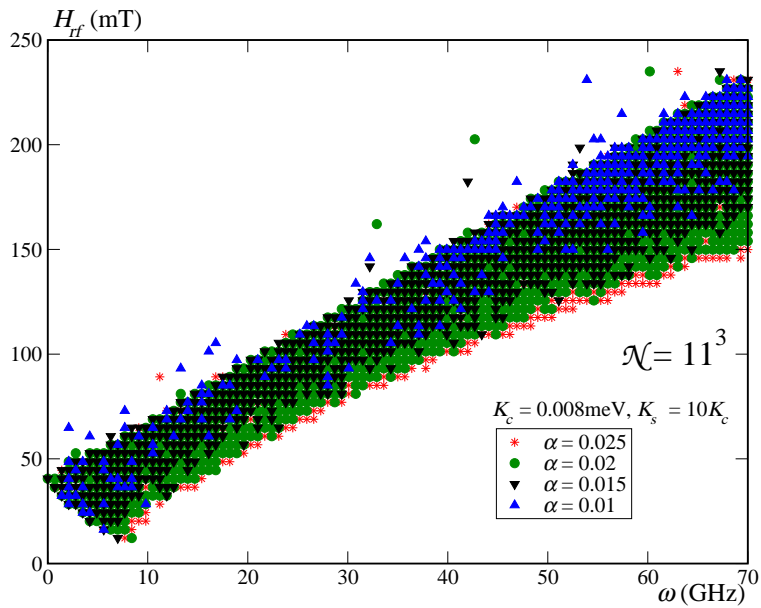} \protect\caption{\label{fig:DampingEffect}Switching phase diagram for different values
of the damping parameter.}
\end{figure}

Now we discuss the effect of the damping parameter $\alpha$ in Eq.
(\ref{eq:LLE}). The present study is concerned with the magnetization
dynamics at very low temperatures. As such, the magnetization switching
can only be achieved by going over the energy barrier. In this case,
the role of the damping term in Eq. (\ref{eq:LLE}) is to drive the
net magnetic moment of the nanomagnet towards the effective field
and thus to push the system into its global energy minimum. In particular,
it does not affect the time trajectory of the net magnetic moment
during its switching process. However, the value of $\alpha$ does
have a bearing on the computing time needed for the system to reach
equilibrium, \emph{i.e.} its global minimum, but does not affect the
global switching diagram as far as the values of $h_{{\rm rf}}$ and
$\varpi$ are concerned. This is indeed confirmed by the results shown
in Fig. \ref{fig:DampingEffect}. Consequently, this study and all
subsequent results have been obtained for a fixed value of $\alpha$,
stated after Eq. (\ref{eq:LLE}).

In general, it is a rather involved task to deal with the dynamics
within a many-spin approach, especially the calculation of relaxation
rates, magnetization reversal, AC susceptibility and so on. Indeed,
it is a formidable task to perform a detailed analysis of the various
critical points (minima, maxima, saddle points) of the energy which
are required for the study of the relaxation processes \citep{lan68prl,lan69ap,kac03epl,kac04jml}.
In the particular case of the present study of the switching $\varpi$-$h_{{\rm rf}}$
phase diagram, it is not so easy to analyze the various switching
mechanisms. As such, it is desirable to replace the many-spin problem
by some effective macroscopic model. Accordingly, it has been shown
\citep{garkac03prl,kacbon06prb,kachkachi07j3m,yanesetal07prb} that,
under some conditions which are quite plausible for today's state-of-the-art
grown nanomagnets, namely of weak surface disorder, one can map this
atomistic approach onto a macroscopic model for the net magnetic moment
$\bm{m}$ defined in Eq. (\ref{eq:Macrospin}), evolving in the effective
potential (H.O.T. = higher-order terms) 
\begin{equation}
\mathcal{E}_{{\rm eff}}=-k_{2}m_{z}^{2}+k_{4}\left(m_{x}^{4}+m_{y}^{4}+m_{z}^{4}\right)+\mbox{H.O.T}.\label{eq:Energy-EOSP}
\end{equation}

We remark in passing that there is a rich literature of works \citep{bertottietal02jap,denisovetal06prl,podbielskietal07prl,wolbac07prl,CaiEtAl_prb13}
that study, in the macrospin approach, the switching of a nanomagnet
assisted by the circularly polarized microwave field in Eq. (\ref{eq:rfField})
and, in some cases, analytical analysis is performed in order to determine
the various stable states of the system and their optimal switching
paths. In the present situation, the effective model in Eq. (\ref{eq:Energy-EOSP}),
as compared to the macrospin studied in the cited works, takes into
account the internal structure and boundary effects in the nanomagnet,
and these lead to the quartic term in the energy potential in Eq.
(\ref{eq:Energy-EOSP}). Because of the latter, highly nonlinear contributions
are brought in and it is thereby difficult, if not impossible, to
carry out similar analytical investigations even at equilibrium.

In the sequel, we will refer to Eqs. (\ref{eq:Energy-EOSP}, \ref{eq:Macrospin})
as the effective-one-spin problem (EOSP) while the many-spin nanomagnet,
described by the set of equations (\ref{eq:HD-Ham}) and (\ref{eq:HamUA}),
will be referred to as the many-spin problem (MSP). We emphasize that
the two leading terms in Eq. (\ref{eq:Energy-EOSP}) have coefficients
$k_{2}$ and $k_{4}$ whose value and sign are functions of the atomistic
parameters ($J,K_{{\rm c}},K_{{\rm s}},$ etc) and of the size and
shape of the nanomagnet \citep{garkac03prl,kachkachi07j3m,Garanin_PhysRevB.98.054427, yanesetal07prb}.
Moreover, we note that both the core and surface may contribute to
$k_{2}$ and $k_{4}$. For example, when the anisotropy in the core
is uniaxial, $k_{2}\simeq k_{{\rm c}}N_{{\rm c}}/\mathcal{N}$, where
$N_{{\rm c}}$ is the number of atoms in the core, see Ref. \citealp{kacbon06prb}.
In fact, even in this case the quartic term appears and is a pure
surface contribution. Regarding the coefficient of the quartic contribution,
for a sphere we have $K_{4}=\kappa K_{s}^{2}/zJ$ (or $k_{4}=\kappa k_{s}^{2}/z$)
where $\kappa$ is a surface integral \citep{garkac03prl} and for
a cube, $k_{4}=\left(1-0.7/\mathcal{N}^{1/3}\right)^{4}k_{s}^{2}/z$
\citep{Garanin_PhysRevB.98.054427}. The details of the conditions
under which this effective model is applicable are discussed in Ref.
\citealp{garkac03prl} and may be summarized as follows: the spin
misalignment (or canting) should not be too strong. Therefore, the
effective model can be built for a nanomagnet whose surface anisotropy
constant $K_{s}$, as compared to the spin-spin exchange coupling
$J$, is small ($k_{s}\lesssim1$). On the other hand, the nanomagnet
size should not be too large for it to be considered as a single magnetic
domain, for a given underlying material.

\begin{figure}[h]
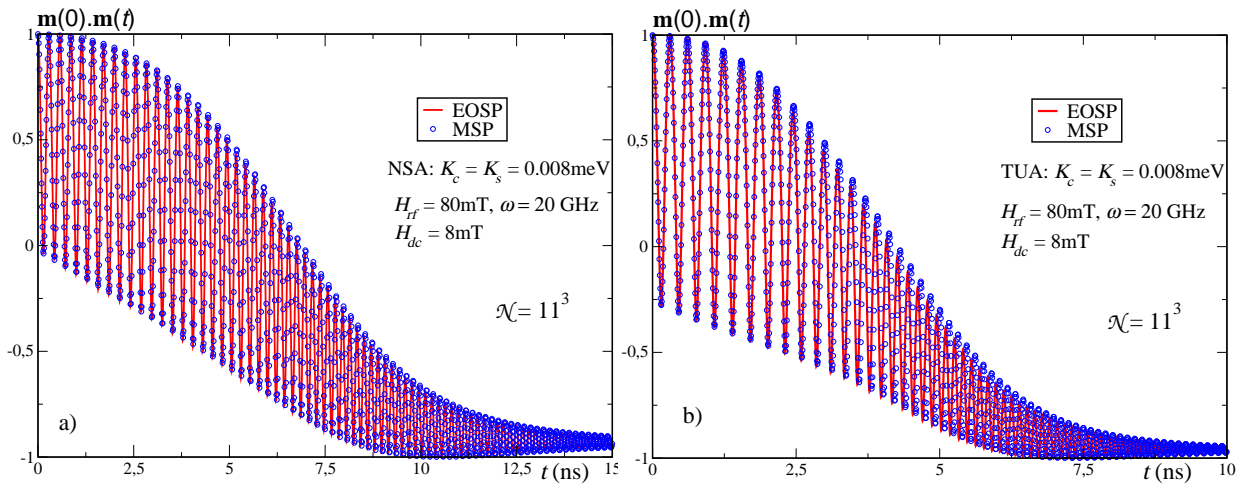

\includegraphics[scale=0.2]{EOSPvsMSP-NSA}\includegraphics[scale=0.2]{EOSPvsMSP-TUA}

\protect\caption{\label{fig:OSPVsMSP}Magnetization temporal evolution for a nanomagnet
of $\mathcal{N}=11^{3}$ spins. (Upper panel) EOSP compared to MSP+NSA
and (lower panel), EOSP compared to MSP and TUA.}
\end{figure}

In Fig. \ref{fig:OSPVsMSP} we plot the time evolution of the net
magnetic moment of a nanomagnet of size $11\times11\times11$ with
the parameters indicated in the legend, for both the NSA (a) and TUA
(b) and . In both situations, we see that for the surface anisotropy
constant considered, the effective model (EOSP) recovers very well
the behavior of the many-spin nanomagnet (MSP). Hence, in the sequel
all results will be presented for the EOSP model, since the corresponding
calculation spares us much CPU time while producing the same results,
as long as the validity conditions for EOSP are met, which will be
the case for all subsequent calculations {[}See discussion below{]}.
Note that when the size and the other parameters of the MSP system
are varied, the coefficients $k_{2}$ and $k_{4}$ of the EOSP model
(\ref{eq:Energy-EOSP}) are accordingly adjusted.

\section{More results and discussion}

\subsection{Temporal evolution}

\begin{figure}[h]
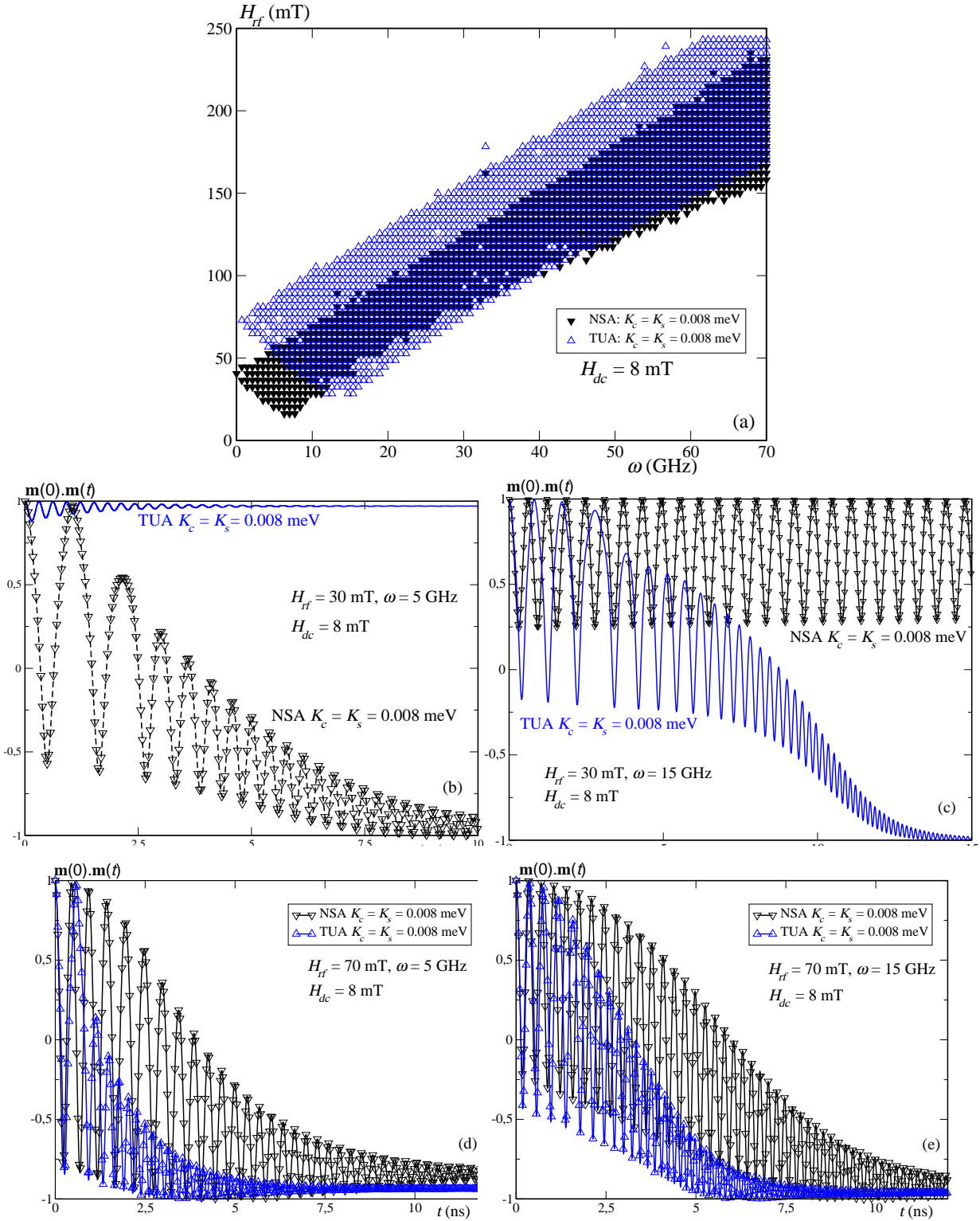

\begin{centering}
\includegraphics[scale=0.25]{PhaseDig-N11-KcKs0008}\\
 \includegraphics[scale=0.21]{mt-TUA-NSA-30mT5GHz}\includegraphics[scale=0.2]{mt-TUA-NSA-30mT15GHz}\\
 \includegraphics[scale=0.2]{mt-TUA-NSA-70mT5GHz}\includegraphics[scale=0.2]{mt-TUA-NSA-70mT15GHz} 
\par\end{centering}
\protect\caption{\label{fig:FBC_NSAVsUni_DCField} Upper panel: phase diagram for TUA
and NSA models. Middle panel: time evolution for $H_{{\rm rf}}=30\,{\rm mT}$
and $\omega=5\ {\rm GHz}$ (left) and $\omega=15\ {\rm GHz}$ (right).
Lower panel: time evolution for $H_{{\rm rf}}=70\,{\rm mT}$ and $\omega=5\ {\rm GHz}$
(left) and $\omega=15\ {\rm GHz}$ (right). In all cases $H_{{\rm dc}}=8\,{\rm mT}$.
$\mathcal{N}=11^{3}$.}
\end{figure}

In Fig. \ref{fig:FBC_NSAVsUni_DCField} we plot the results for a
nanomagnet of size $\mathcal{N}=11\times11\times11$. The upper panel
is the ($\omega,H_{{\rm rf}}$) phase diagram for TUA and NSA. In
both models, the value of the anisotropy constant is the same for
all spins within the nanomagnet. The physical parameters used are
given in the caption and legends. The result in Fig. \ref{fig:FBC_NSAVsUni_DCField}(a)
shows that the NSA model with a nonuniform distribution of the anisotropy
easy axes allows for switching in a region with smaller amplitude
and frequency (left lower corner of the diagram) than the TUA model.
On the other hand, the latter allows for a wider domain of switching
field intensities, though of higher values. Next, the graph in Fig.
\ref{fig:FBC_NSAVsUni_DCField}(b) shows that, for a given amplitude
and frequency of $h_{{\rm rf}}$, switching is obtained with NSA and
not with TUA. Again, this means that the misalignment induced by the
Néel surface anisotropy favors switching. This result, however, depends
on frequency as is shown by the graphs in Fig. \ref{fig:FBC_NSAVsUni_DCField}(b)
and (c). Now, comparing the graphs in Figs. \ref{fig:FBC_NSAVsUni_DCField}(b)
and (d), we see that switching is achieved within TUA but at the expense
of a higher rf field amplitude. Under this field, and higher frequencies,
switching is achieved in both models, as witnessed by the graph in
Fig. \ref{fig:FBC_NSAVsUni_DCField}(e). 

The general explanation of these observations resides in the anisotropy
configuration within the two models, NSA and TUA, for a box-shaped
nanomagnet. For the TUA model, the anisotropy is textured with all
easy axes pointing in the $z$ (vertical) direction. So, on the horizontal
facets, NSA and TUA have the same easy axis\citep{kacbon06prb}. However,
on the other four facets and along the edges, NSA has a perpendicular
effective anisotropy direction. As such, in the TUA model the effective
anisotropy field is stronger (more rigid) and thereby the magnetization
switching requires higher external fields and/or higher frequencies.

\subsection{Effect of surface anisotropy ($K_{s}$)}

\begin{figure*}[t]
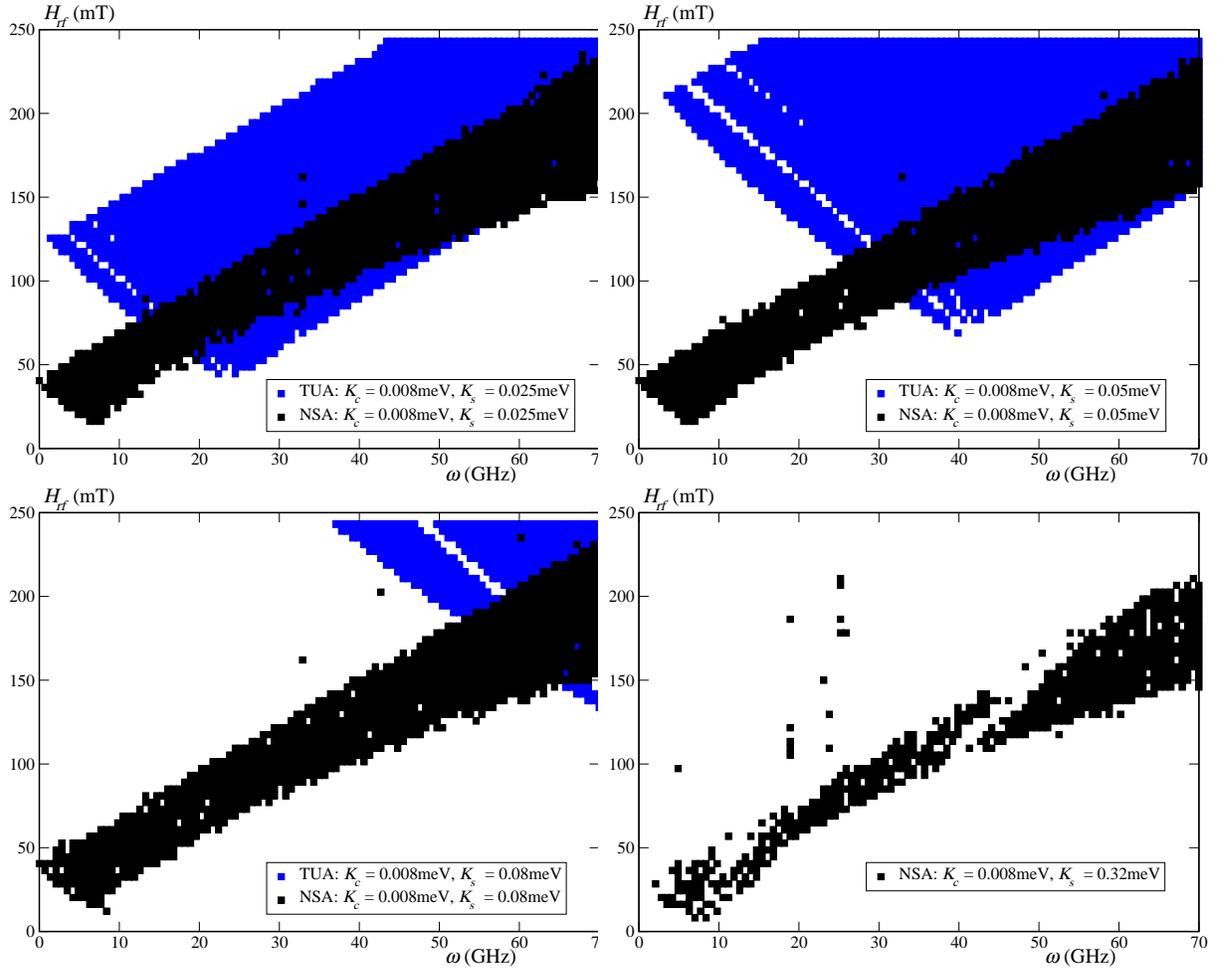

\includegraphics[scale=0.2]{KsEffect-0025}\includegraphics[scale=0.2]{KsEffect-005}\\
 \includegraphics[scale=0.2]{KsEffect-008}\includegraphics[scale=0.2]{KsEffect-032}

\protect\caption{\label{fig:PhaseDiagram-11-vKs}Phase diagram for TUA and NSA with
increasing (downwards) surface anisotropy constant $K_{s}$. $H_{{\rm dc}}=8\,{\rm mT}$.
$\mathcal{N}=11^{3}$.}
\end{figure*}

The first three phase diagrams in Fig. \ref{fig:PhaseDiagram-11-vKs}
are obtained for TUA and NSA models with the same (fixed) anisotropy
constant in the core and increasing anisotropy constant at the surface.
Comparing the three diagrams, we first observe that the TUA model
renders a phase diagram that shifts to the right along the diagonal,
\emph{i.e.} towards higher amplitudes and frequencies of the rf field.
This is due to the fact that as $K_{s}$ increases it makes the system
with textured anisotropy (\emph{i.e.} TUA) more rigid. On the other
hand, the NSA seems to be less sensitive to the increase of $K_{s}$,
apart from a small shift to lower amplitudes of $h_{{\rm rf}}$. In
fact, once the latter has reached the sufficient value for inducing
a nonuniform spin configuration, the switching mechanism remains the
same. However, as $K_{s}$ is further increased, the phase diagram
shows more ``holes'' which means that switching is no longer possible
for all field amplitudes and frequencies. This is shown by the last
phase diagram in Fig. \ref{fig:PhaseDiagram-11-vKs}. In fact, when
surface anisotropy is strong enough (but not too strong) as to induce
spin misalignment, several generations of spin waves are created with
different amplitudes and frequencies \citep{Yukalov_prb05,YukalovEtAl_prb08,garkacrey08epl,garkac09prb},
but only those modes that are at resonance or whose frequencies match
the rf field frequency do participate in the switching mechanism \citep{podbielskietal07prl,wolbac07prl}.
On the other hand, the results in the lower right panel of Fig. \ref{fig:PhaseDiagram-11-vKs}
show that the effective model (EOSP) reaches the limits of its validity
regarding the parameters $K_{s}$. Indeed, in Refs. \citep{kacbon06prb,yanesetal07prb},
it was checked that the validity limit on $k_{s}=K_{s}/J$ is about
$0.25$ for a simple cubic lattice and $0.35$ for a face-centered
cubic lattice.

Summing up, it is clearly demonstrated that surface effects, as exemplified
by the NSA model, open up switching channels for relatively low rf
fields intensities and frequencies, as compared to the textured-anisotropy
model TUA. This is seen in all phase diagrams in Fig. \ref{fig:PhaseDiagram-11-vKs}
and in subsequent ones. Indeed, they all exhibit a switching area
that stretches to the left lower corner, though they remain narrower
at higher intensities and frequencies.

In general, in a nanomagnet with nonuniform anisotropy, transverse
spin waves are excited which are not necessarily spin waves in an
equilibrium state. These transverse spin fluctuations trigger spin
motion which may grow up into a coherent dynamics and ultimately induce
magnetization switching\citep{Yukalov_prb05,YukalovEtAl_prb08,garkacrey08epl,garkac09prb}.

\subsection{Size effect}

\begin{figure*}
\includegraphics[scale=0.2]{SizeEffect-N7} \includegraphics[scale=0.2]{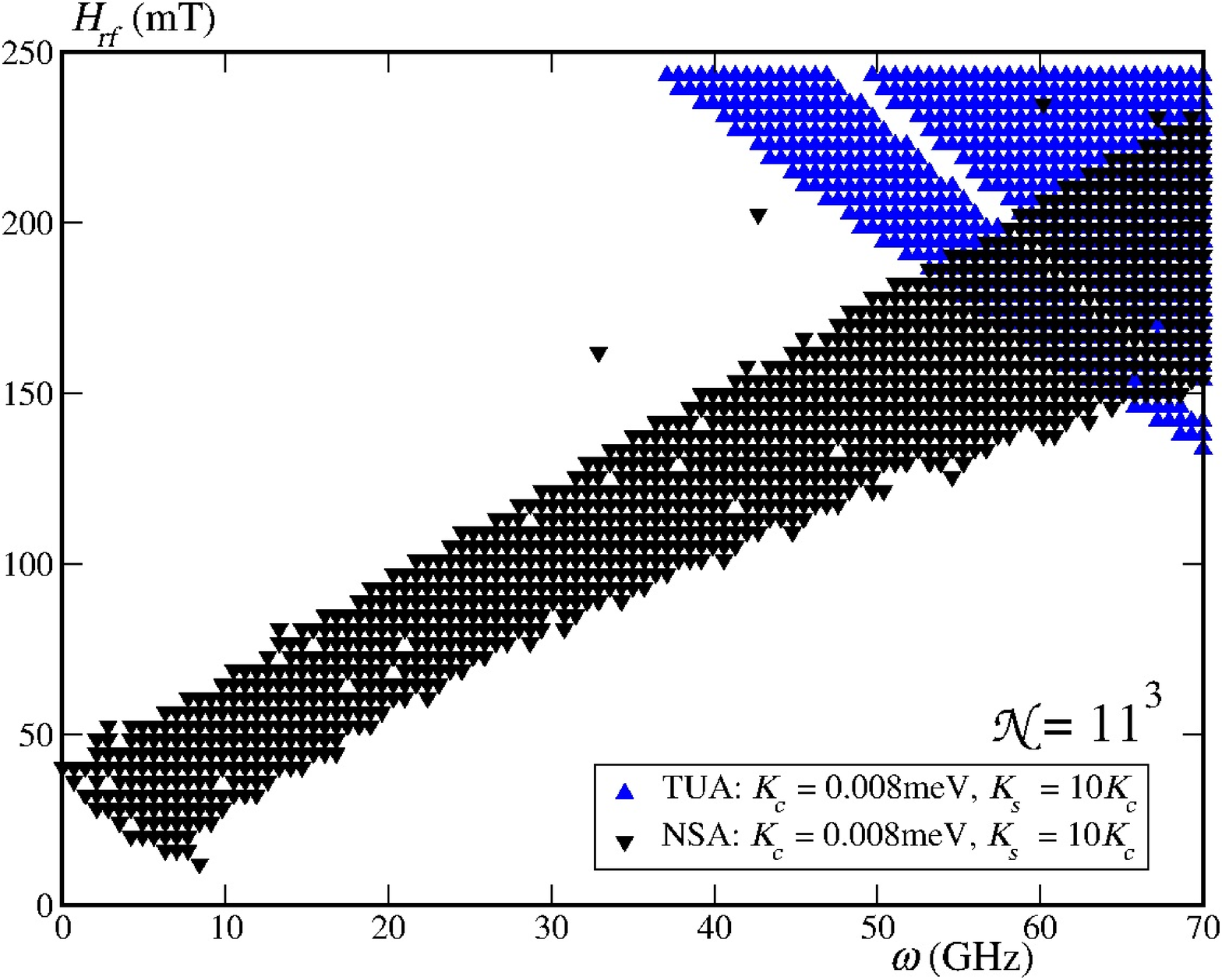}\\

\includegraphics[scale=0.2]{SizeEffect-N15}\includegraphics[scale=0.2]{SizeEffect-N25}\\
\includegraphics[scale=0.2]{SizeEffect-N50}\includegraphics[scale=0.2]{SizeEffect-N100}

\protect\caption{\label{fig:PhaseDiagram-vN}Phase diagram for the model NSA with varying
size $\mathcal{N}=7^{3},11^{3},15^{3},25^{3},50^{3},100^{3}$.}
\end{figure*}

In Fig. \ref{fig:PhaseDiagram-vN} we plot the $\varpi$-$h_{{\rm rf}}$
phase diagram for both the NSA and TUA nanomagnets of varying size
(cube side), $N=9,11,15,25,50,100$ which corresponds to a ratio of
surface-to-total number of spins equal to $N_{{\rm st}}=64\%,45\%,35\%,22\%,12\%,6\%$,
respectively. The anisotropy constants are chosen so that $K_{s}/K_{c}=10$. 

First, we see that, in what regards the TUA model, as the size increases
($N_{{\rm st}}$ decreases) the phase diagram extends towards lower
amplitudes and frequencies of the rf field. For example, for the size
$N=7$, corresponding to $N_{{\rm st}}=64\%$, \emph{i.e.} with a
large surface contribution (in number of spins and anisotropy strength),
the phase diagram of TUA ``disappears'' which means that switching
in this case would require much higher values of the rf field amplitude
and frequency. On the other hand, upon increasing the size, the surface
contribution decreases in terms of spins number, and thereby the core
effective anisotropy becomes relatively stronger than that on the
surface and, as such, low-frequency and small-field dynamics takes
over. This is illustrated in Fig. \ref{fig:PhaseDiagram-vN} by the
results for size $\mathcal{N}\geq11$. 

The situation for the NSA model is somewhat reversed. First, for all
these sizes and the corresponding surface contributions, magnetization
switching is achieved for all rf fields (amplitude and frequency)
in the explored ranges. Moreover, the NSA phase diagram widens in
the lower left corner of the diagram, \emph{i.e.} for lower amplitudes
and frequencies of the rf field when $N_{{\rm st}}$ decreases. This
means that, as was shown by the results in Fig. \ref{fig:PhaseDiagram-11-vKs},
when the contribution of surface anisotropy dominates, there are only
some specific modes that contribute to switching. 

As can be seen in Fig. \ref{fig:PhaseDiagram-vN}, at very large system
sizes (\emph{e.g. }$50,100$), the surface contribution becomes negligible
and the two models tend to (asymptotically) yield the same switching
diagram, since the black diagram widens and the blue one shrinks.
This is mainly due to the fact that as $N_{{\rm st}}\rightarrow0$,
the effective anisotropy in the two models becomes the same.

\section{Conclusion}

We have studied the effect of surface anisotropy, and the spin misalignment
it induces, on the magnetization switching in a many-spin nanomagnet
subjected to an rf magnetic field, on top of a DC magnetic field.
Our study is based on the numerical solution of the damped Landau-Lifshitz
equation for exchange-coupled atomic spins.

We have demonstrated that surface effects, as exemplified by Néel's
model for surface atoms, open up switching channels with relatively
low rf fields intensities and frequencies, as compared to the model
with parallel easy axes, or the so-called macrospin model. In general,
these favorable switching conditions depend on the size and shape
of the nanomagnet and, of course, on the underlying magnetic properties
and energy parameters. However, thanks to today's state of the art
facilities for growing well defined nanomagnets, it is possible to
meet most of these conditions and achieve an optimized magnetization
switching in these nanoscale systems. On the other hand, most of the
systems available today are ensembles of interacting nanomagnets with
controlled separation and spatial organization. The present study
has been devoted to a single nanomagnet in order to investigate the
(intrinsic) surface effects on the magnetization switching mechanisms.
It will be of fundamental and practical interest to extend this study
to interacting nanomagnets and to compare with measurements on assemblies
of different concentrations.

\bibliographystyle{apsrev4-2}
%

\end{document}